\def\e{{\rm e}}
\def\del{\partial}
\def\half{{1\over2}}

\def\abs#1{{\left|{#1}\right|}}
\def\vev#1{\langle #1 \rangle}
\def\del{\partial}
\def\dslash{\del\kern-0.55em\raise 0.14ex\hbox{/}}

\def\rough#1{\raise.3ex\hbox{$#1$\kern-.75em\lower1ex\hbox{$\sim$}}}
\newcommand{\PRD}[3]{{\it Phys. Rev.} {\bf D{#1}} (19{#3}) {#2}}
\newcommand{\PRDM}[3]{{\it Phys. Rev.} {\bf D{#1}} (20{#3}) {#2}}

\newcommand{\NPB}[3]{{\it Nucl. Phys.} {\bf B{#1}} {#2} (19{#3})}
\newcommand{\NPBM}[3]{{\it Nucl. Phys.} {\bf B{#1}} (20{#2}) {#3}}
\newcommand{\PLB}[3]{{\it Phys. Lett.} {\bf {#1}B} (19{#3}) {#2}}
\newcommand{\PLBM}[3]{{\it Phys. Lett.} {\bf {#1}B} (20{#3}) {#2}}

\newcommand{\PTP}[3]{{\it Prog. Theor. Phys.} {\bf {#1}} (19{#3}) {#2}}

\newcommand{\ANN}[3]{{\it Ann. Phys. (N.Y.)} {\bf {#1}}, {#2} (19{#3})}

\newcommand{\ZP}[3]{{\it Zeit. Phys.} {\bf C{#1}}, {#2} (19{#3})}
\newcommand{\MPL}[3]{{\it Mod. Phys. Lett.} {\bf A{#1}} (19{#3}) {#2}}
\newcommand{\MPLM}[3]{{\it Mod. Phys. Lett.} {\bf A{#1}} (20{#3}) {#2}}

\newcommand{\jhep}[3]{{\it J. High Energy Phys.}{\bf {#1}}, {#2} (20{#3})}

\newcommand{\hepph}[1]{{\tt hep-ph/#1}}
\newcommand{\hmu}{\hat\mu}

\documentclass[12pt]{article}
\usepackage{graphicx}
\begin{document}
\baselineskip=18pt
%%%%%%%%%%%%%%%%%%%%%%%%%%%%
\begin{titlepage}
%%%%% PREPRINT NUMBERS %%%%%%
\begin{flushright}
RIKEN-TH-42\\
OU-HET-526/2005 \\
hep-th/0505066
\end{flushright}
%%%%%%%%%%%%%%%%%%% TITLE %%%%%%%%%%%%%%%%%%
\begin{center}{\Large\bf Aspects of Phase Transition \\
\vspace*{3mm}
in Gauge-Higgs Unification 
at Finite Temperature}
\end{center}
%%%%%%%%%%%%%%%% AUTHORS %%%%%%%%%%%%%%%%%%%%%%%
\vspace{1cm}
\begin{center}
Nobuhito Maru$^{(a)}$
\footnote{E-mail: maru@riken.jp},
Kazunori Takenaga$^{(b)}$
\footnote{E-mail: takenaga@het.phys.sci.osaka-u.ac.jp}
\end{center}
%%%%%%%%%%%%%%%%%%%%%%% AFFILIATION %%%%%%%%%%%%
\vspace{0.2cm}
\begin{center}
%\small
${}^{(a)}$ {\it Theoretical Physics Laboratory, RIKEN, 
Wako, Saitama 351-0198, Japan}
\\[0.2cm]
${}^{(b)}$ {\it Department of Physics, Osaka University, 
Toyonaka, Osaka 560-0043, Japan}
%%%%%
%${}^{(d)}$ {\it School of Theoretical Physics,
%Dublin Institute for Advanced Studies, \\
%10 Burlington Road, Dublin 4, Ireland}
%%%%%%%
\end{center}
%%%%%%%%%%%%%%%%%% ABSTRACT %%%%%%%%%%%%%%%
\vspace{1cm}
\begin{abstract}
We study the phase transition in gauge-Higgs unification 
at finite temperature. 
%The first order phase transition is observed 
%for the gauge symmetry breaking patterns we consider.
In particular, we obtain the strong first order electroweak 
phase transition for a simple matter content 
yielding the correct order of Higgs mass at zero temperature. 
Two stage phase transition is found 
for a particular matter content, 
which is the strong first order at each stage. 
%At each stage, 
%the phase transition is also strongly the first order. 
We further study supersymmetric gauge models 
with the Scherk-Schwarz supersymmetry breaking. 
We again observe the first order electroweak phase transition 
%for simple matter contents 
and 
%point out that 
multi stage phase transition. 
%(four stages in our case) 
%phase transitions occur for a certain matter content.
\end{abstract}
\end{titlepage}
%%%%%%%%%%
%\tableofcontents
%%%%%%%%%%%%
\newpage
%%%%%%%%%%%%%%%% INTRODUCTION %%%%%%%%%%%%%%%
\section{Introduction}
There has been paid much attention to the scenario of 
gauge-Higgs unification \cite{gaugehiggs1}\cite{gaugehiggs2}
%%%
%The scenario of gauge-Higgs unification has been paid much
%attention 
%%%%
for the possibility to solve the gauge hierarchy problem 
without supersymmetry (SUSY). 
%%%
%introducing new symmetry. 
%%%
In this scenario, Higgs fields are regarded as extra 
components in the original gauge field. 
The higher dimensional local gauge invariance ensures that the Higgs 
field is strictly massless. If spatial coordinates are 
compactified, then the extra component gauge field behaves as the 
scalar field at low energies. And the scalar field  can
develop the vacuum expectation values (VEV) through the dynamics
of the Wilson line phases, which is 
called as the Hosotani mechanism \cite{hosotani}.
\par
%%%%%%%%%%%
The dynamics of the Wilson line phases has been studied 
from various points of view \cite{gaugehiggs3}. 
In particular, an attempt to identify the Wilson line degrees 
of freedom with the Higgs scalar in the standard model (or
in the minimal SUSY standard model) has been explored. 
Since the mass of the Higgs scalar in the gauge-Higgs unification is
calculable, insensitive to the ultraviolet physics\footnote{In
Gravity-Gauge-Higgs unification, the mass of Higgs scalar 
identified with the extra components of the metric is also calculable
and insensitive to the ultraviolet physics. 
However, there are subtleties
in diagrammatic calculations, see ref.\cite{maru}.} and 
is generated by quantum
corrections through the 
Coleman-Weinberg mechanism \cite{cw}, 
so that it is natural to expect that the Higgs mass is light, at most, 
the order of the gauge boson mass. The general analyses have been
made in refs.\cite{haba}\cite{hytmodel} and 
pointed out that the Higgs mass can be 
as heavy as possible to satisfy the present experimental 
lower bound of the Higgs mass. The scenario is one of the 
promising possibility beyond the standard model 
even though there are many issues that should be studied 
such as the fermion mass hierarchy problem.
\par
%%%%%%%%%%% 
The dynamics of the Wilson line phases is essentially 
the Coleman-Weinberg mechanism, 
so that the transition at finite temperature is expected 
to be the first order \cite{finitec}. 
In fact, 
the authors of ref.\cite{finitee} extensively studied 
the nature of the phase transition 
of the gauge-Higgs unification at finite temperature and 
investigated the origin of the first order phase transition. 
They have explicitly demonstrated the first order phase transition 
in certain models \cite{silve} and shown that 
the lighter Higgs mass gives us the stronger first order phase transition. 
The strong first order phase transition is favored 
from a point of view of the scenario of 
the electroweak baryogenesis \cite{baryon}. 
\par
%%%%%%%%%%
In this paper we study the gauge-Higgs unification 
at finite temperature in other models, 
including supersymmetric gauge models, 
which are different from the ones studied in ref.\cite{finitee}. 
In our model, there is no need to introduce massive 
fields, and the matter content is very simple. 
We do not consider matter fields belonging to the higher dimensional 
representation under the gauge group except the adjoint one. 
In this matter content, the mass of the Higgs scalar 
can be consistent with the experimental lower bound. 
We obtain that the phase transition is the first order and 
find that the transition is so strong as to satisfy 
the necessary condition for the electroweak baryogenesis. 
We also observe the tendency that heavier the
Higgs mass becomes, weaker the first order phase transition is. 
We calculate the critical temperature at which the two degenerate
vacua appear and estimate how strong the phase transition is 
in our models.  
\par
%%%%%%%%%
We also find an interesting phenomena for a certain matter content 
even in the the gauge-Higgs unification at finite temperature. 
There are two stage phase transitions, 
which has also been observed in the four dimensional
SUSY model at finite temperature \cite{finited}. In our 
model at the first critical temperature, 
the vacuum with the $SU(2)\times U(1)$ gauge symmetry is degenerate 
with that respecting the $U(1)^{\prime}\times U(1)$ gauge symmetry. 
Decreasing the temperature further, at the second critical temperature, 
the vacuum with the $U(1)^{\prime}\times U(1)$ gauge 
symmetry is degenerate with the vacuum with the $U(1)_{em}$ gauge symmetry. 
Both phase transitions are the first order at each phase transition. 
We also analyze the strength of the phase transition 
at each critical temperature.  
\par
%%%%%%%%%%%
We also study SUSY models with the Scherk-Schwarz 
SUSY breaking \cite{ss} at finite temperature. 
We again find the phase transition to be very strong first order, 
which is favored from the electroweak baryogenesis. 
The relevant quantity to measure 
the strength of the phase transition is also evaluated. 
We also find multi (four in our case) phase transition 
in the SUSY model with a particular matter content. 
Except for the first phase transition 
where the vacua with the $SU(2)\times U(1)$
gauge symmetry and the $U(1)_{em}$ gauge 
symmetry are degenerate at the critical temperature, 
all of these vacua respect the $U(1)_{em}$ gauge symmetry 
are degenerate in the rest of the three phase transitions. 
\par
%%%%%%%
%We are further interested in
%the effect of the chemical potential on the phase transition
%at finite temperature. We obtain the expression for the effective
%potential with the chemical potential. 
%\par
%%%%%%%%%
In the next section, 
we investigate the behavior of the effective potential 
for the Wilson line phases at finite temperature in our model 
and show that the phase transition is the first order. 
Then, we estimate the critical temperature and 
the strength of the phase transition. 
%which is a relevant quantity for the
%discusssion of the electroweak baryogenesis. 
The phase transition is found to be very strong and 
provides us an interesting possibility for the electroweak baryogenesis. 
In section $3$,
we study the case of SUSY gauge models with 
the Scherk-Schwarz SUSY breaking and study the effective potential 
for the Wilson line phases at finite temperature. 
For a particular matter content, 
we find multi phase transition, 
four stages phase transitions in our case, 
in which the phase transition is the first order. 
The final section is devoted to conclusions and discussions. 
%%%%%%%%%%%%
\section{Non supersymmetric gauge models}
%%%%
%\section{Models and phase transition}
%%%
Let us consider gauge theory coupled with matter at finite
temperature in $D$ dimensions, where one of the space coordinate is
compactified on an orbifold, $S^1/Z_2$. 
The space time is regarded as an 
$S^1\times M^{D-2}\times S^1/Z_2$, 
where the Euclidean time direction $\tau$ is effectively compactified 
on a circle, $S^1$ whose periodicity is given 
by $\beta\equiv T^{-1}$, where $T$ stands for the temperature, 
the radius of the other $S^1$ and its coordinate 
are denoted by $R$ and $y$, respectively. We decompose the $D$
dimensional gauge field $A_{\hmu}(\hmu = 0,1,D-1)$ 
as $A_{\hmu}=(A_{\tau}, A_{\mu}, A_y)$.  
\par
%%%%%%%%
When one studies the gauge theory on the space time 
with boundaries, one needs to specify boundary conditions of fields 
for the compactified direction. 
The boundary condition for the Euclidean time direction 
is determined by quantum statistics, 
so that one uniquely assigns the (anti-)periodic boundary condition 
for (fermions)bosons,\footnote{Let us note that 
the ghost field must obey the periodic boundary condition 
for the Euclidean time direction \cite{hatakugo}.} 
while for the orbifold $S^1/Z_2$, 
we must specify the boundary condition 
on the two fixed points of the orbifold, $y=0$ and $y=\pi R$ 
in addition to the $S^1$ direction. 
We define that
\begin{eqnarray}
A_{\hmu}(x, y + 2\pi R) &=&
U A_{\hmu}(x, y ) \, U^\dagger ,
\label{shiki1}
\\
\pmatrix{A_\mu \cr A_y \cr} (x, z_i - y) &=&
P_i \pmatrix{A_\mu \cr - A_y \cr} (x, z_i + y) \, 
P_i^\dagger  \quad (i = 0, 1),
\label{shiki2}
\end{eqnarray}
where $U^\dagger = U^{-1}, P_i^\dagger = P_i= P_i^{-1}$ and $z_0=0,
z_1=\pi R$. The minus sign for $A_y$ is needed to preserve the gauge
invariance under these transformations. Since a transformation 
$\pi R+y \rightarrow \pi R -y$ must be the same as a transformation 
$\pi R +y \rightarrow -(\pi R + y)\rightarrow \pi R -y$, we obtain that
\begin{equation}
U = P_1 P_0
\label{shiki3}
\end{equation}  
Hereafter, we consider $P_i$ is more fundamental quantity than $U$.
\par
%%%%%%%%%%%
The orbifolding boundary conditions $P_i$ determine 
the gauge symmetry breaking patterns at the tree level. 
%%%
%The gauge symmetry breaking patterns depend on the choice of $P_i$. 
%%%%
In this paper, we start with an $SU(3)$ gauge group 
and choose $P_0=P_1=\mbox{diag.}(1,1,-1)$. 
Since the zero mode field for $A_{\mu}^a$ is given by the generators 
of $SU(3), \lambda^{a=1,2,3,8}$ commuting 
with $P_{0,1}$, the $SU(3)$ breaks down to 
the $SU(2)\times U(1)$ gauge group. 
We are interested in the gauge symmetry breaking patterns 
of $SU(2)\times U(1)$ at finite temperature 
after taking quantum corrections into account, that is, 
the phase transition through the dynamics of the Wilson line phases 
at finite temperature.
\par
%%%%%%%%%
The Wilson line phases for the choice of $P_{0,1}=\mbox{diag}(1,1,-1)$
is given by the zero mode field $A_y^b$ associated with 
the $SU(3)$ generators, $\lambda^{b=4, 5, 6, 7}$, which anticommutes 
with $P_{0,1}$. Then, the zero mode,
\begin{equation}
\Phi \equiv \sqrt{2\pi R}\pmatrix{A_y^4 -i A_y^5 \cr A_y^6 -i A_y^7 \cr} 
\label{shiki4}
\end{equation}
transforms as an $SU(2)$ doublet, so that we can regard $\Phi$ as a
Higgs doublet. We see that the degrees of freedom of the
Higgs doublet is embedded in the zero mode part of $A_y$.
This is the idea of the gauge-Higgs unification. 
\par
%%%%%%%%%%%%%
We paramerize $\vev{A_y}$, utilizing 
the $SU(2)\times U(1)$ degrees of freedom, as
\begin{equation} 
\vev{A_y} ={a\over{g_4 R}}{\lambda^6\over 2}
\equiv A_y^{6(0)}{\lambda^6\over 2}
\label{shiki5}
\end{equation}
where $g_4\equiv g/\sqrt{2\pi R}$ and $a$ is a real
parameter. The parameter $a$ is related with the Wilson line phases, 
\begin{eqnarray}
W &=& {\cal P} \mbox{exp} \left(ig \oint_{S^1}dy A_y \right)
= \pmatrix{
1 & 0 & 0 \cr 
0 & \cos(\pi a)  & i\sin(\pi a) \cr
0 & i\sin(\pi a) & \cos(\pi a) }\qquad (a~~\mbox{mod}~2)
\nonumber\\
&=& \left\{
\begin{array}{lll}
SU(2)\times U(1) & \mbox{for} & a=0, \\[0.1cm]
U(1)^{\prime}\times U(1)& \mbox{for} & a=1 \\[0.1cm]
U(1)_{em} & \mbox{for} & \mbox{otherwise}
\end{array}\right.
\label{shiki6}
\end{eqnarray}
We observe that, depending on the values of $a$, 
the gauge symmetry breaking patterns are different, 
and in order to obtain the correct pattern of 
electroweak symmetry breaking, 
one needs the fractional values of $a$. As we will see later, the 
values of $a$ is determined dynamically  
through the dynamics of the Wilson line phases by minimizing the
effective potential for the phase $a$. 
%We are interested in the phase
%transition for the gauge symmetry breaking patterns.
\par
%%%%%%%%%%
%\subsection{Non supersymmetric gauge models}
%%%%%%%%
Now, let us introduce fermions 
belonging to the adjoint (fundamental) 
representation under the gauge group $SU(3)$ and denote their flavor number
by $N_{adj}^{(\pm)} (N_{fd}^{(\pm)})$, where 
the sign $(\pm)$ stands for the intrinsic parity defined 
in refs.\cite{hhhk}\cite{haba}. 
Similarly, $N_{adj}^{(\pm)s}(N_{fd}^{(\pm)s})$ means 
the flavor number for bosons belonging to 
%the adjoint (fundamental) 
corresponding representations 
%under the $SU(3)$ with the intrinsic parity $(\pm)$.
\par
%%%%%%%%%%%%
Following the standard prescription \cite{hosotani}, 
it is straightforward to calculate the zero
temperature part of the effective potential for the Wilson line 
phase $a$ (corresponding to the background field (\ref{shiki5})),
\begin{eqnarray} 
V_{eff}^{T=0}(a) &=&
{\Gamma(D/2) \over{{\pi^{D/2}(2\pi R)^D}}} 
\sum_{n=1}^{\infty}{1 \over n^D} \nonumber\\
&\times & 
\Biggl[
\left(-(D-2)+(r2^{[D/2]}) N_{adj}^{(+)}-N_{adj}^{(+)s}\right)
\left(\cos[2\pi na] + 2\cos[\pi n a] \right) 
\nonumber
\\
&+&\left((r2^{[D/2]}) N_{adj}^{(-)}-N_{adj}^{(-)s}\right) 
\left( \cos[2\pi n(a - \half)] +2 \cos[\pi n(a-1)] \right) 
\nonumber
\\
&+& \left((r2^{[D/2]}) N_{fd}^{(+)}-2N_{fd}^{(+)s} \right)
\cos[\pi n a] + 
\left((r2^{[D/2]}) N_{fd}^{(-)}- 2N_{fd}^{(-)s}\right) 
\cos[\pi n (a-1)]
\Biggr],\nonumber \\
\label{shiki7}
\end{eqnarray}
where we have assumed the adjoint scalar is real 
and $r$ is $\half ({1\over 4})$ for Majorana (Majorana-Weyl fermion). 
We have ignored the irrelevant $a$-independent terms in Eq.(\ref{shiki7}). 
\par
%%%%%%%%%%%
Let us briefly present the main results of the reference \cite{haba}.
We set $D=5$, which is implicitly assumed throughout this paper.
For the flavor number chosen as  
\begin{eqnarray}
N_{adj}^{(+)s} &=& 0,~~N_{adj}^{(-)s} = 0,~~N_{adj}^{(+)} = 2,
~~N_{adj}^{(-)}=0, \nonumber\\
N_{fd}^{(+)} &=& 0,~~N_{fd}^{(-)} = 8,~~N_{fd}^{(+)s}=4,
~~N_{fd}^{(-)s} = 2, 
\label{shiki8}
\end{eqnarray}
the minimum of the effective potential (\ref{shiki7}) is 
dynamically determined to be $a_0\simeq 0.0583$, 
so that the $SU(2)\times U(1)$ correctly breaks down to $U(1)_{em}$. 
The Higgs mass, which is obtained by 
the second derivative of the effective potential 
evaluated at the minimum, is calculated as 
\begin{equation}
m_H^2 \equiv {{\del^2 V_{eff}^{T=0}} 
\over{\del {A_y^{6(0)}}^2}}\Bigg |_{a=a_0}
=(gR)^2 {{\del^2 V_{eff}^{T=0}} \over{\del a^2}} \Bigg|_{a=a_0}
\simeq \left(g_4^2 \times 129~~\mbox{GeV}\right)^2. 
\label{shiki9}
\end{equation} 
In the scenario of the gauge-Higgs unification, 
we notice an important relation,
\begin{equation}
\vev{\Phi}={a_0 \over{g_4R}} = v \sim 246~\mbox{GeV},
\label{shiki10}
\end{equation}
which has been used in Eq.(\ref{shiki9}).
\par
%%%%%%%%%%%
Let us now consider the finite temperature part 
and study the phase transition of the model with the matter content
(\ref{shiki8}). 
Following the standard prescription of 
the finite temperature field theories \cite{finitea}\cite{finiteb}, 
the finite temperature part of the effective potential is 
obtained as
\begin{eqnarray}
V_{eff}^{T \neq 0}(a)
&=&
{{2\Gamma(D/2)} \over{{\pi^{D/2}(2\pi R)^D}}} 
\sum_{l=1}^{\infty}
\sum_{n=1}^{\infty}
{1 \over{\left[(2\pi R n)^2 + l^2/T^2 \right]^{D/2}}} \nonumber\\
&\times& 
\Biggl[
\left(-(D-2)+(r2^{[D/2]}) N_{adj}^{(+)} (-1)^l- N_{adj}^{(+)s} \right)
\left(\cos[2\pi na] + 2 \cos[\pi n a] \right)
\nonumber\\
&+&\left((r2^{[D/2]}) N_{fd}^{(+)} (-1)^l-2 N_{fd}^{(+)s}\right) 
\cos[\pi n a] 
\nonumber\\
&+& \left((r2^{[D/2]}) N_{adj}^{(-)} (-1)^l-N_{adj}^{(-)s} \right)
\left( \cos[2\pi n(a - \half)] + 2\cos[\pi n(a-1)] \right)
\nonumber\\
&+& \left((r2^{[D/2]}) N_{fd}^{(-)} (-1)^l-2N_{fd}^{(-)s}\right)
 \cos[\pi n (a-1)] \Biggr],
\label{shiki11}
\end{eqnarray}
where we have ignored the irrelevant constant. The total effective 
potential we study is given by
\begin{equation}
V_{eff}(a) = V_{eff}^{T=0}(a) + V_{eff}^{T \neq 0}(a).
\label{shiki12}
\end{equation}
It is useful to introduce a dimensionless quantity 
$z \equiv RT$ for numerical studies. We also note that the total
effective potential is invariant under $a\rightarrow -a$ and
$a\rightarrow 2-a$, which implies that it is enough to consider 
the region given by $0\leq a \leq 1$.
\par
\begin{figure}
\begin{center}
\includegraphics[width=9cm,height=9cm,keepaspectratio]{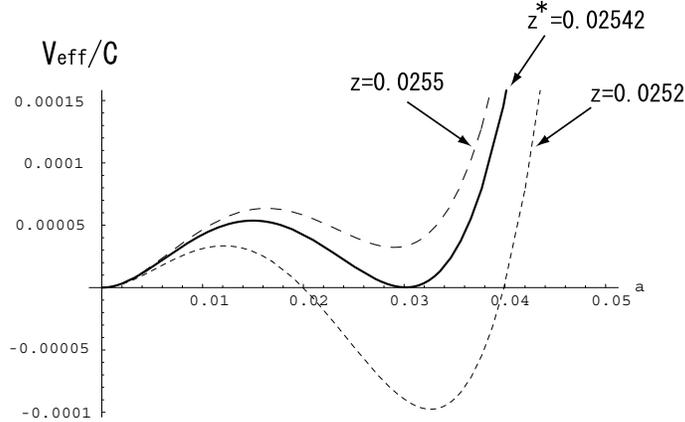}
%%%%%%%%%%
%\begin{picture}(0,0)
%\put(-40,90){$z=0.0252$}
%\put(-50,100){$\nwarrow$}
%\put(-50,150){$z^{\star}=0.02542$}
%\put(-61,140){$\swarrow$}
%\put(-155,110){$z=0.0255$}
%\put(-100,100){$\searrow$}
%\end{picture}
%%%%%%%%%%%%%%
\end{center}
\caption{The behavior of the effective potential $V_{eff}(a)/C$ 
for a non-SUSY model with the matter content (\ref{shiki8}) 
is drawn with respect to $z\equiv RT$, 
where $C\equiv \Gamma(5/2)/\pi^{5/2}(2\pi R)^5$.}
\label{zu1}
\end{figure}

%%%%%%%%%%% 
For the matter content given by Eq.(\ref{shiki8}), we study the 
behavior of the effective potential with respect to the 
various values of $z$. 
Figure \ref{zu1} tells us that the phase transition is the first order. 
%%%%%%%%%%%%%%
The critical temperature $T_c$, at which the two degenerate vacua appear, is 
numerically obtained as
\begin{equation}
z^{\star} \equiv  RT_c \simeq 0.02542,
\label{shiki13}
\end{equation}
which gives the critical temperature 
\begin{equation}
T_c \simeq {1 \over R} \times 0.02542 = 
\left({{g_4 v} \over a_0} \right) \times 0.0254
\simeq g_4 \times 107.1~~\mbox{GeV}
\label{shiki14}
\end{equation}
where we have used the relation (\ref{shiki10}). 
\par
%%%%%%%%%%%%%%
When one discusses the electroweak baryogenesis, the strength 
of the phase transition at finite temperature is crucial to satisfy the
famous Zakharov's conditions. Namely, the strong first order phases
transition is required to decouple the sphaleron process to leave 
the generated baryon. 
The relevant quantity we examine is $v(T_c)/T_c$ 
and should be larger than the unity, 
which is calculated as
\begin{equation}
{v(T_c) \over T_c} = \left({{a_0^{T_c}} \over{g_4 R}} \right)
{1 \over T_c} = {1\over g_4}{{a_0^{T_c}} \over z^{*}} 
\simeq {1 \over g_4} \times 1.19 > 1,
%
%~~\mbox{for}~~g_4\sim O(1),
%
\label{shiki15}
\end{equation}
where $a_0^{T_c}$ 
%at which the two degenerate vacua appear,
is numerically calculated as 
\begin{equation}
a_0^{T\neq 0} \simeq 0.0302.
\label{shiki16}
\end{equation}
The phase transition for the model with the matter content (\ref{shiki8})
through the dynamics of the Wilson line phases (the Hosotani mechanism) 
is found to be the strong first order as seen from Eq.(\ref{shiki15}) 
%the transition is strongly first order.
\footnote{We take $g_4\sim O(1)$ in
this paper as in refs. \cite{haba}, \cite{hytmodel}.}. 
Hence it may be possible 
to make use of this first order phase transition 
for the electroweak baryogenesis in the scenario of gauge-Higgs unification. 
One should observe that the light Higgs mass (small $g_4$) tends to make 
the phase transition stronger, which is consistent with the 
results obtained in ref.\cite{finitee}.
\par
%%%%%%%%%%%%%
Next, we study the phase transition for 
other gauge symmetry breaking patterns determined by the values of $a$.
This study is important to understand aspects and nature of the 
phase transition in the gauge-Higgs unification at finite temperature.
\par
%%%%%%%%%%%%%
Let us choose the following extremely simplified matter content as 
\begin{eqnarray}
N_{adj}^{(+)s} &=& 0,~~N_{adj}^{(-)s}=0,~~N_{adj}^{(+)}=0,~~N_{adj}^{(-)}=0, 
\nonumber\\
N_{fd}^{(+)} &=& 3,~~N_{fd}^{(-)}=0,~~N_{fd}^{(+)s}=0,~~N_{fd}^{(-)s}=0,
\label{shiki17}
\end{eqnarray}
where the system corresponds to the gauge theory coupled to only 
the fundamental fermions. By numerical studies, we find the VEV, 
$a_0$ at the minimum of the effective potential (\ref{shiki7}) 
is given by
\begin{equation}
a_0 = 1.0.
\label{shiki18}
\end{equation}
From Eq.(\ref{shiki6}), we see that the unbroken gauge symmetry 
is $U(1)^{\prime} \times U(1)$, and the neutral gauge 
boson corresponding to the $Z$ boson in the
standard model remains massless. 
The mass of the scalar that respects the $U(1)^{\prime} \times U(1)$ 
gauge symmetry is obtained as before,
%
%taking the second derivative of $V_{eff}^{T=0}$ at the minimum, as 
%
\begin{equation}
m_H^2\simeq \left(g_4^2 \times 24.03~~\mbox{GeV}\right)^2. 
\label{shiki19}
\end{equation}
\par
\begin{figure}
\begin{center}
\includegraphics[width=9cm, height=9cm,keepaspectratio]{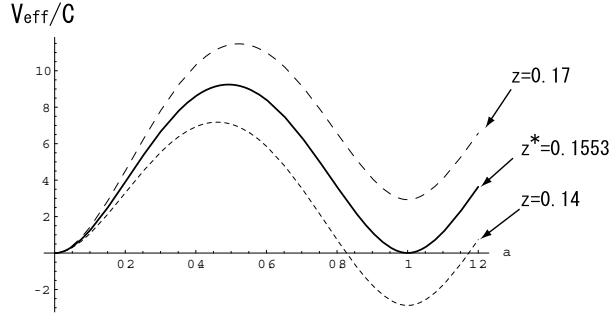}
%%%%%%%%%%
%\begin{picture}(0,0)
%\put(-8,52){$z=0.14$}
%\put(-20,43){$\swarrow$}
%\put(-8,82){$z^{\star}=0.1553$}
%\put(-20,73){$\swarrow$}
%\put(-8,110){$z=0.17$}
%\put(-20,100){$\swarrow$}
%\end{picture}
%%%%%%%%%%%%%%
\end{center}
\caption{The behavior of the effective potential $V_{eff}(a)/C$ 
for a non-SUSY model with the matter content (\ref{shiki17}) is drawn 
with respect to $z\equiv RT$. 
%where $C\equiv \Gamma(5/2)/\pi^{5/2}(2\pi R)^5$ 
}
\label{zu2}
\end{figure}

%%%%%%%%%%%%%%%%%%%%
Taking the temperature effect into account, 
we can see again from Fig.~\ref{zu2} 
that the phase transition is the first order. 
The behavior of the effective potential is quite different from 
that of the previous example.
%%%%% 
%(see figure $2$). 
%%%%%%%%%%%%%%
%%%%%%%%%%%%%%
Contrary to the previous case, 
the local minimum for $T \geq T_c$ is always located as
$a_0^{T\geq T_c}=1.0$.
%%%
%even at nonzero temperature. 
%%%%
The critical temperature, at which the vacua with 
the $SU(2)\times U(1)$ gauge symmetry and 
the $U(1)^{\prime}\times U(1)$ gauge symmetry are degenerate, 
is given by
\begin{equation}
T_c \simeq {1 \over R} \times 0.1553 = \left({{g_4 v}\over
a_0}\right) 
\times 0.1553 \simeq g_4 \times 38.20~~\mbox{GeV}. 
\label{shiki20}
\end{equation}
We calculate the relevant quantity to see how strong the phase
transition is as
\begin{equation}
{v(T_c) \over T_c} = \left({{a_0^{T_c}} \over{g_4 R}}\right){1 \over T_c}
= {1 \over g_4}{{a_0^{T_c}} \over z^{*}} 
\simeq  {1 \over g_4} \times 6.392 > 1. 
%
% > 1~~\mbox{for}~~g_4 \sim O(1).
%
\label{shiki21}
\end{equation}
We observe again the strong first order phase transition, and 
the mass of the scalar respecting the $U(1)^{\prime}\times U(1)$ gauge
symmetry is very light, as shown in Eq.(\ref{shiki19}). 
\par
%%%%%%%%%%%%%
Let us finally consider the case 
where we introduce only the adjoint fermions 
instead of the fundamental one,
\begin{eqnarray}
N_{adj}^{(+)s} &=& 0,~~N_{adj}^{(-)s}=0,~~N_{adj}^{(+)}=2,~~N_{adj}^{(-)}=0, 
\nonumber\\
N_{fd}^{(+)} &=&0,~~
N_{fd}^{(-)}=0,~~N_{fd}^{(+)s}=0,~~N_{fd}^{(-)s}=0.
\label{shiki22}
\end{eqnarray}
At zero temperature, the gauge symmetry correctly 
breaks down to $U(1)_{em}$, and the VEV is given by 
\begin{equation}
a_0 = 0.6667,
\label{shiki23}
\end{equation}
for which the Higgs mass is calculated as  
\begin{equation}
m_H^2 \simeq  \left(g_4^2 \times 32.41~~\mbox{GeV}\right)^2.
\label{shiki24}
\end{equation}
The Higgs boson is too light to be consistent 
with the present experimental lower bound. 
This, however, implies the very strong first order phase 
transition, as pointed out in ref.\cite{finitee}. In 
order to see this property, 
let us turn on the temperature effects. 
We unexpectedly observe interesting phase transition
patterns for the matter content (\ref{shiki22}). 
There are two stage phase transitions 
as the temperature decreases.\footnote{The two
phase transition patterns have been observed 
in the next to SUSY standard model ref.\cite{finited}. 
%for a certain supersymmetric models. 
In the next section, 
we will also see a multi phase transition 
in other SUSY models.} 
\par
\begin{figure}
\begin{center}
\includegraphics[width=9cm, height=9cm,keepaspectratio]{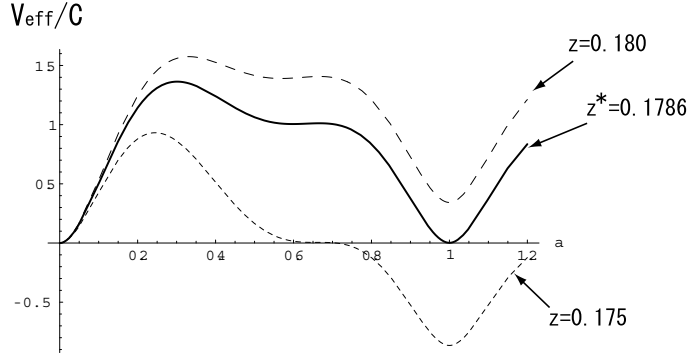}
\end{center}
\caption{The behavior of the effective potential $V_{eff}(a)/C$ 
for a non-SUSY model with the matter content (\ref{shiki22}) is drawn 
with respect to $z\equiv RT$. 
%%%
%where $C\equiv \Gamma(5/2)/\pi^{5/2}(2\pi R)^5$. 
%%%%
The first phase transition of the two stage phase transitions is shown.}
%%%%%%%%%%
%\begin{picture}(0,0)
%\put(348,94){$z=0.175$}
%\put(335,100){$\nwarrow$}
%\put(355,180){$z=0.1786$}
%\put(340,170){$\swarrow$}
%\put(355,200){$z=0.180$}
%\put(340,190){$\swarrow$}
%\end{picture}
%%%%%%%%%%%%%%
\label{zu3}
\end{figure}

%%%%%%%%%%%
The first transition occurs at $z_1^{\star}\simeq 0.1786$, 
so that the critical temperature is 
\begin{equation}
T_{1c} \simeq {1 \over R} \times 0.1786 \simeq g_4 \times
65.866~~\mbox{GeV},
\label{shiki25}
\end{equation}
where we have used (\ref{shiki10}). At this critical temperature,
there appear two degenerate vacua, that is, 
the vacua with the $SU(2)\times U(1)$ gauge symmetry and 
the $U(1)^{\prime}\times U(1)$ gauge 
symmetry (see figure \ref{zu3}).
%%%%%%%%%%%%%%
%%%%%%%%%%%%%%%%%%%%%%%%
\begin{figure}
\begin{center}
\includegraphics[width=9cm, height=9cm,keepaspectratio]{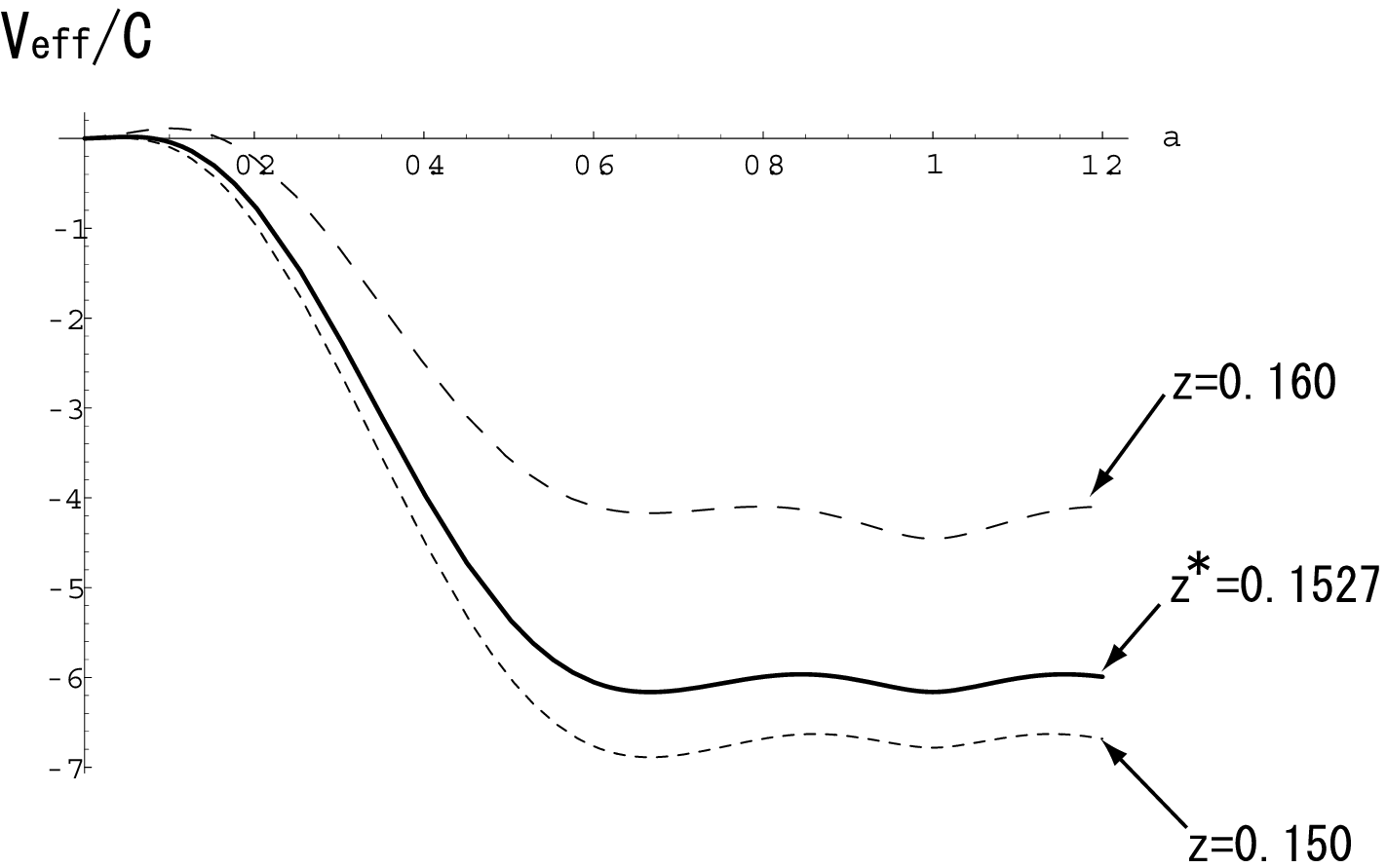}
\end{center}
\caption{The behavior of the effective potential $V_{eff}(a)/C$ 
for a non-SUSY model with the matter content (\ref{shiki22}) is drawn 
with respect to $z\equiv RT$. 
%%%% 
%where $C\equiv \Gamma(5/2)/\pi^{5/2}(2\pi R)^5$ 
%%%
The second phase transition of the two stage phase transition 
is shown.}
%%%%%%%%%%%%%
%\begin{picture}(0,0)
%\put(348,65){$z=0.150$}
%\put(335,70){$\nwarrow$}
%\put(355,105){$z^{\star}=0.1527$}
%\put(340,95){$\swarrow$}
%\put(355,140){$z=0.160$}
%\put(340,130){$\swarrow$}
%\end{picture}
%%%%%%%%%%%%%%
\label{zu4}
\end{figure}
%%%%%%%%%%%%%%%
The phase transition is the first order and 
the strength of the phase transition is obtained as 
\begin{equation}
{v(T_{1c}) \over T_{1c}} = 
\left({{a_0^{T_{1c}}} \over{g_4 R}}\right){1\over T_{1c}}
= {1 \over g_4}{{a_0^{T_{1c}}} \over z_1^{*}} 
\simeq  {1\over g_4} \times 5.599 > 1,
%
%> 1~~\mbox{for}~~g_4\sim O(1),
%
\label{shiki26}
\end{equation}
where
\begin{equation}
a_0^{T_{1c}} = 1.0.
\label{shiki27}
\end{equation}
Let us note that, as before, the local minimum of the effective
potential even at nonzero temperature is always
located at $a_0^{T=T_{1c}}=1.0$. 
\par
%%%%%%%%%%
As the temperature decreases further, the second 
phase transition occurs at
\begin{equation} 
T_{2c} \simeq {1 \over R} \times 0.1527 \simeq g_4 \times
56.088~~\mbox{GeV},
\label{shiki28}
\end{equation}
at which the vacua with the $U(1)^{\prime}\times U(1)$ gauge symmetry 
and the $U(1)_{em}$ gauge symmetry are 
degenerate (see figure \ref{zu4}). We obtain the 
strength of the phase transition in this case as
\begin{equation} 
{v(T_{2c}) \over T_{2c}} ={1\over g_4}{{\Delta a_0^T}\over
z_2^{\star}}\simeq {1 \over g_4} \times 2.183 > 1, 
%
%> 1~~\mbox{for}~~g_4 \sim O(1).
%
\label{shiki29}
\end{equation}
where $\Delta a_0^T\simeq  1.0-0.666667$. Let us 
also note that the local minima of 
the effective potential is
again always located at $a_0^{T<T_{2c}}\simeq 0.6667$,
which is the global minimum of the zero temperature part of the
effective potential. This type of two stage phase transitions 
has never been observed in the scenario of the gauge-Higgs unification at
finite temperature, and it is very interesting to consider physical 
applications of this type of the phase transition. 
We will see multi phase transition in the case of supersymmetric 
gauge models in the next section.  
\par
%%%%%%%%
It has been known that fermions and/or the fermion belonging to 
the higher dimensional represenation, in general, tend to weaken the
first order phase transition \cite{finitea}. Here we confirm 
the statement is true 
from Eqs.(\ref{shiki21}), (\ref{shiki26}) and (\ref{shiki29}). 
\par
%%%%%%%%%%
Before closing this section, 
we would like to mention the difference between the work by 
Panico and Serone \cite{finitee} and ours. In their paper, a 
similar $SU(3)$ gauge model of gauge-Higgs unification 
at finite temperature was investigated and the first order electroweak 
phase transitions were obtained. 
In their model, in order to obtain the viable Higgs mass, 
two options were discussed, one is to introduce the bulk fields 
in the higher rank representation under $SU(3)$ gauge group 
(i.e. symmetric tensor), 
the other is to introduce localized gauge kinetic terms. 
However, the strength of the first order phase transition is at most 
$v(T_c)/T_c \simeq 0.13(0.7)$ for the former(latter) option. 
On the other hand, as discussed in the text, 
there is no need to introduce bulk fields 
in the higher rank representation except for the adjoint one in our model. 
Furthermore, the strength of the phase transition is stronger 
(larger than the unity). 
Thus, we can conclude that our model is more favorable from the viewpoint 
of the application to the electroweak baryogenesis. 
%%%%%%%%%%
\section{Supersymmetric gauge models}
%%%%%%%%%%%%%%%%
In this section, let us proceed to 
supersymmetric gauge models. We first study the
phase transition for the 
the model considered in ref.\cite{hytmodel}. As before, the degrees of
freedom for the Wilson
line phase is given by the parameter $a$ in
Eq.(\ref{shiki6}). The effective potential for the phase 
at $T=0$ is given by
\begin{eqnarray}
V_{eff}^{T=0}(a) &=&
{{4\times \Gamma(5/2)} \over{{\pi^{5/2}(2\pi R)^5}}} 
\sum_{n=1}^{\infty}{1 \over n^5}\left(1-\cos[2\pi n\beta]\right) \nonumber\\
&\times & 
\Biggl[
\left(-1+N_{adj}^{(+)}\right)
\left(\cos[2\pi na] + 2\cos[\pi n a] \right) 
\nonumber
\\
&+&N_{adj}^{(-)}\left( \cos[2\pi n(a - \half)] +2 \cos[\pi n(a-1)] \right) 
\nonumber
\\
&+& N_{fd}^{(+)}\cos[\pi n a] + N_{fd}^{(-)} \cos[\pi n (a-1)]
\Biggr],
\label{shiki30}
\end{eqnarray}
%%%%%%%%%%%
%V_{SUSY}^{T=0} &=& \frac{4\Gamma(5/2)}{\pi^{5/2}(2 \pi R)^5}
%\sum_{n=1}^{\infty}
%\Biggl[
%-\left(J^{(+)}[2a,\beta,n] + 2J^{(+)}[a,\beta,n]\right)\nonumber\\
%&&+N_{adj}^{(+)}\left(J^{(+)}[2a,\beta,n] +
%2J^{(+)}[a,\beta,n]\right)
%\nonumber \\
%&&+N_{adj}^{(-)}\left(J^{(-)}[2a,\beta,n] + 2J^{(-)}[a,\beta,n]\right)
%\nonumber \\
%&&+N_{fd}^{(+)} J^{(+)}[a,\beta,n] +N_{fd}^{(-)} J^{(-)}[a,\beta,n]
%\Biggr], 
%\end{eqnarray}
%%%%%%%%%%%%%%%%
where $\beta$ is the Scherk-Schwarz (SS) parameter which explicitly
breaks supersymmmetry by the boundary condition for the $S^1$ direction 
for the gaugino and squarks in the supermultiplet \cite{ss}. As we can see,
$\beta=0~(\mbox{mod}~1)$ yields the vanishing effective potential due to
the recovery of the original supersymmetry.
\par
%%%%
%\begin{eqnarray}
%J^{(+)}[a,\beta,n] &\equiv& \frac{1}{n^5} (1-{\rm cos}(2 \pi n \beta)) 
%{\rm cos}(\pi n a), 
%\\
%J^{(-)}[a,\beta,n] &\equiv& \frac{1}{n^5} (1-{\rm cos}(2 \pi n \beta))
%{\rm cos}(\pi n (a-1)).
%
%\end{eqnarray}
%%%%%%%
Let us choose the flavor number and the SS parameter as 
\begin{equation}
N_{adj}^{(+)}=2,~~N_{adj}^{(-)}=2,~~N_{fd}^{(+)}
=0,~~N_{fd}^{(-)}=2,~~\beta=0.14.
\label{shiki31}
\end{equation}
In ref.\cite{hytmodel}, it has been shown that for the choice (\ref{shiki31}),
the electroweak gauge symmetry is correctly broken, and the VEV is
given by
\begin{equation}
a_0\simeq 0.0379.
\label{shiki32}
\end{equation}
The Higgs mass is obtained as 
\begin{equation}
m_H^2 
%%%%
%\equiv {{\del^2 V_{SUSY}^{T=0}} 
%\over{\del {A_y^{6(0)}}^2}}\Bigg |_{a=a_0}
%=(gR)^2 {{\del^2 V_{SUSY}^{T=0}} \over{\del a^2}} \Bigg|_{a=a_0}
%%%%
\simeq \left(g_4^2 \times 130~~\mbox{GeV} \right)^2.
\label{shiki33}
\end{equation} 
\par
%%%%%%%
We next consider the finite temperature effects. 
The effective potential at $T\neq 0$ is calculated as
%%%%%%%%%%%
\begin{eqnarray}
V_{SUSY}^{T \ne 0}(a) &=& -\frac{2\times 4\Gamma(5/2)}
{\pi^{5/2}(2 \pi R)^5}  
\times \sum_{l=1}^{\infty} \sum_{n=1}^{\infty} 
\frac{1}{[(2 \pi R n)^2 + l^2/T^2]^{5/2}} \nonumber \\
&\times& \Biggl[
\left(\tilde{J}^{(+)}[2a,\beta,n] + 2\tilde{J}^{(+)}[a,\beta,n] \right) 
+N_{adj}^{(+)}
\left(\hat{J}^{(+)}[2a,\beta,n] + 2\hat{J}^{(+)}[a,\beta,n]\right)
\nonumber \\
%&&+N_{adj}^{(+)}
%\left(\hat{J}^{(+)}[2a,\beta,n] + 2\hat{J}^{(+)}[a,\beta,n]\right) 
%\nonumber \\
&+& 
N_{adj}^{(-)}\left(\hat{J}^{(-)}[2a,\beta,n]+2\hat{J}^{(-)}[a,\beta,n]\right) 
%\nonumber \\
%&& 
+N_{fd}^{(+)}\hat{J}^{(+)}[a,\beta,n] 
+N_{fd}^{(-)}\hat{J}^{(-)}[a,\beta,n]
\Biggr], \nonumber \\
\label{shiki34}
\end{eqnarray}
where we have defined
\begin{eqnarray}
\tilde{J}^{(+)}[a,\beta,n] &\equiv& (1-(-1)^l {\rm cos}(2 \pi n \beta))
{\rm cos}(\pi n a), \label{shiki35}\\
\hat{J}^{(+)}[a,\beta,n] &\equiv& ({\rm cos}(2 \pi n \beta)-(-1)^l)
{\rm cos}(\pi n a), \label{shiki36}\\
\hat{J}^{(-)}[a,\beta,n] &\equiv& ({\rm cos}(2 \pi n \beta)-(-1)^l)
{\rm cos}(\pi n (a-1)).
 \label{shiki37}
\end{eqnarray}
We observe that even if the SS parameter is zero we obtain the
nonvanishing effective potential at the finite temperature due to
the difference of statistics between bosons and fermions. 
The total effective potential we study is, then, given by 
\begin{equation}
V_{SUSY}^{tot}(a)=V_{SUSY}^{T=0}(a)+V_{SUSY}^{T\neq 0}(a).
\label{shiki38}
\end{equation}
Let us note that $V_{SUSY}^{tot}(a)$ is invariant under
$a\rightarrow -a$ and $a\rightarrow 2-a$, which means that it is 
enough to consider the region $0\leq a \leq 1$ \footnote{As
for the SS parameter, it is enough to consider the region 
$0\leq \beta\leq 1$ because of the 
invariance of $V_{SUSY}^{tot}$ under $\beta\rightarrow 2-\beta$.}. 
\par
\begin{figure}
\begin{center}
\includegraphics[width=9cm, height=9cm,keepaspectratio]{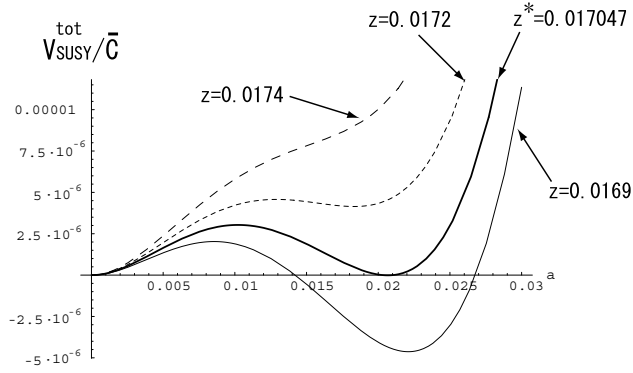}
\end{center}
\caption{The behavior of the effective potential
$V_{SUSY}^{tot}(a)/{\bar C}$ for a SUSY model with the matter content 
(\ref{shiki31}) is drawn with respect to $z\equiv RT$, where ${\bar C}
\equiv 4\Gamma(5/2)/\pi^{5/2}(2\pi R)^5$.}
\label{zu5}
%%%%%%%%%%%%%
%\begin{picture}(0,0)
%\put(190,180){$z=0.0174$}
%\put(200,170){$\searrow$}
%\put(248,220){$z=0.0172$}
%\put(300,210){$\searrow$}
%\put(340,220){$z^{\star}=0.017047$}
%\put(328,210){$\swarrow$}
%\put(353,185){$z=0.0169$}
%\put(340,190){$\nwarrow$}
%\end{picture}
%%%%%%%%%
\end{figure}

%%%%%%%%%%%%%%%%%%
In Figure \ref{zu5}, 
we show the behavior of $V_{SUSY}^{tot}(a)$ for 
various values of $z\equiv RT$. 
%%%%%%%%%%%%%
%%%%%%%%%%%%%%
Clearly, one can see that the phase
transition is the first order, and the critical temperature, at
which the two degenerate vacua appear, is given by
\begin{equation}
T_c \simeq {1\over R} \times 0.017047 \simeq g_4 \times 110.7~~\mbox{GeV},
\label{shiki39}
\end{equation}
where we have used (\ref{shiki10}). The relevant quantity to measure the 
strength of the phase transition is calculated as 
\begin{equation}
{v(T_{c}) \over T_{c}} = {1\over g_4}
{{a_0^{T=T_c}}\over {z^{\star}}}
\simeq {1 \over g_4} \times 1.212 > 1, 
%
%> 1~~\mbox{for}~~g_4 \sim O(1).
%
\label{shiki40}
\end{equation}
where we have used the values $a_0^{T=T_c}=0.0206688$.
The phase transition is so strong as to decouple the sphaleron 
process to leave the created baryon number after the
electroweak phase transition.
\par
%%%%%%%%%%%
Next let us choose the matter content with only fundamental 
representations as 
\begin{equation}
N_{adj}^{(+)}=0,~~N_{adj}^{(-)}=0,~~N_{fd}^{(+)}
=3,~~N_{fd}^{(-)}=0,~~\beta=0.14.
\label{shiki41}
\end{equation}
The minimum of the effective potential at zero temperature
is given by the VEV 
\begin{equation}
a_0=1.0,
\label{shiki42}
\end{equation}
for which the $SU(2)\times U(1)$ gauge symmetry is broken to
$U(1)^{\prime}\times U(1)$. The mass of the scalar that respects 
the residual
gauge symmetry is calculated as
\begin{equation}
m_H^2 
%\equiv {{\del^2 V_{SUSY}^{T=0}} 
%\over{\del {A_y^{6(0)}}^2}}\Bigg |_{a=a_0}
%=(gR)^2 {{\del^2 V_{SUSY}^{T=0}} \over{\del a^2}} \Bigg|_{a=a_0}
\simeq \left(g_4^2 \times 18.03~~\mbox{GeV} \right)^2.
\label{shiki43}
\end{equation}
%If we consider the finite temperature effect, the behavior of the 
%effective potential (\ref{shiki37}) with respect to $z\equiv RT$
%is depicted in Figure $5$.
%%%%%%%%%%%
\begin{figure}
\begin{center}
\includegraphics[width=9cm, height=9cm,keepaspectratio]{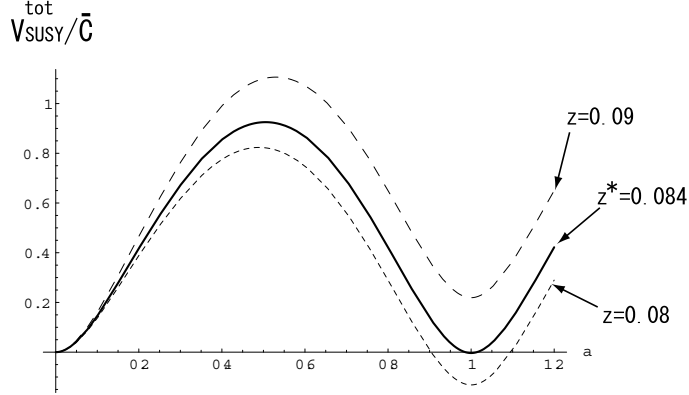}
\end{center}
\caption{The behavior of the effective 
potential $V_{eff}^{tot}(a)/{\bar C}$ for a SUSY model 
with the matter content (\ref{shiki41}) is drawn with respect to $z\equiv RT$. 
%where ${\bar C}\equiv 4\Gamma(5/2)/\pi^{5/2}(2\pi R)^5$.
}
%%%%%%%%%%%%%
%\begin{picture}(0,0)
%\put(350,100){$z=0.08$}
%\put(340,110){$\nwarrow$}
%\put(355,148){$z^{\star}=0.084$}
%\put(340,138){$\swarrow$}
%\put(355,170){$z=0.09$}
%\put(340,160){$\swarrow$}
%\end{picture}
%%%%%%%%%%%%%%
\label{zu6}
\end{figure}
\par
%%%%%%%%%%
We see from Fig.\ref{zu5} that the phase transitions is the first order, 
and the critical temperature is calculated as
\begin{equation}
T_c \simeq {1\over R} \times 0.084 \simeq g_4 \times 20.66~~\mbox{GeV}.
\label{shiki44}
\end{equation}
Similar to the nonsupersymmetric case with the matter 
content (\ref{shiki17}), the local minimum of the effective
potential before $T\geq T_c$ is always given 
by $a_0^{T\geq T_c}=1.0$. We next estimate the relevant quantity
for the electroweak baryogenesis,
\begin{equation}
{v(T_{c}) \over T_{c}} = {1\over g_4}
{{a_0^{T=T_c}}\over {z^{\star}}}
\simeq {1 \over g_4} \times 11.90 > 1, 
\label{shiki45}
\end{equation}
where we have used $a_0^{T=T_c}=1.0$ in Eq.(\ref{shiki45}).  
The phase transition is obviously found to be the first order.
\par
\begin{figure}
\begin{center}
\includegraphics[width=8cm, height=8cm,keepaspectratio]{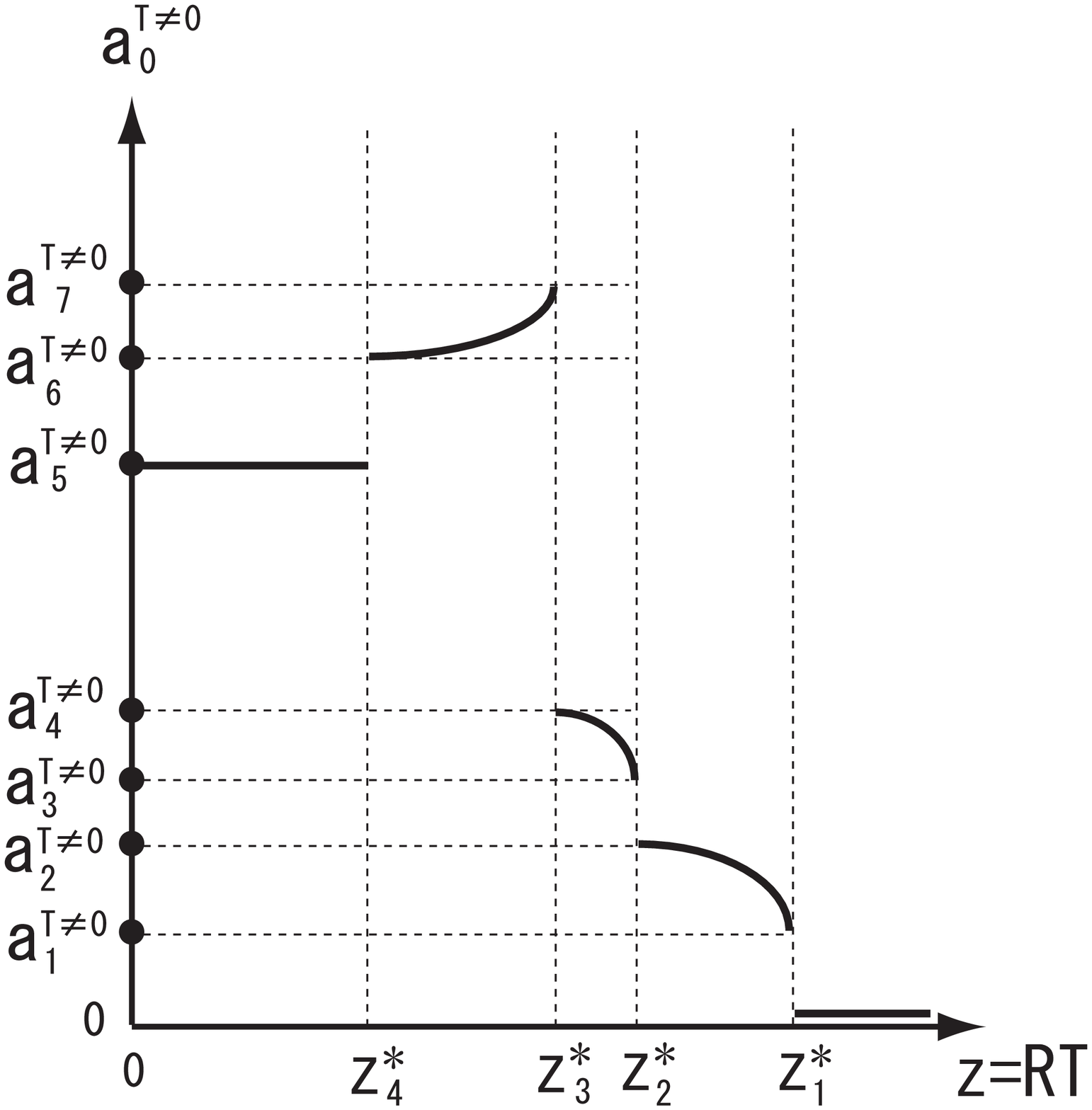}
\end{center}
\caption{The behavior of the order parameter $a_0^{T\neq 0}$ 
for a SUSY model with the matter 
content (\ref{shiki46}) is drawn with respect to $z\equiv RT$.}
\label{zu7}
\end{figure}

%%%%%%%%%%%
Finally, let us consider the matter content with only 
adjoint representations,
\begin{equation}
N_{adj}^{(+)}=2,~~N_{adj}^{(-)}=0,~~N_{fd}^{(+)}=0,
~~N_{fd}^{(-)}=0,~~\beta=0.14.
\label{shiki46}
\end{equation}
The minimum of the effective potential $V_{SUSY}^{T=0}$
is given by the VEV 
\begin{equation}
a_0^{T=0}\simeq 0.6667,
\label{shiki47}
\end{equation}
and thus we have the correct pattern of electroweak symmetry breaking. 
The Higgs mass in this case is found to be very light as 
\begin{equation}
m_H^2 
%\equiv {{\del^2 V_{SUSY}^{T=0}} 
%\over{\del {A_y^{6(0)}}^2}}\Bigg |_{a=a_0}
%=(gR)^2 {{\del^2 V_{SUSY}^{T=0}} \over{\del a^2}} \Bigg|_{a=a_0}
\simeq \left(g_4^2 \times 17.91~~\mbox{GeV} \right)^2.
\label{shiki48}
\end{equation}
The phase structure at finite temperature is unexpectedly rich. 
There is a multi phase transition, in fact, we have
four stages phase transitions for the matter 
content (\ref{shiki46}). It is too complicated to draw the
behavior of the effective potential. Instead, let
us show the behavior of the VEV, $a_0^{T\neq 0}$ in figure \ref{zu7}.
%%%%%%%   
%%%%%%%%
Let us note that at the first phase transition, 
the vacua with the $SU(2)\times U(1)$ gauge symmetry and 
the $U(1)_{em}$ gauge symmetry is 
degenerate, but at the other three phase 
transitions, the vacua with the $U(1)_{em}$ are 
degenerate each other. 
\par
%%%%%%%%%
The critical temperature is obtained
as
\begin{equation}
T_{ic}(i=1\sim 4)(\mbox{GeV})={1\over R}\times 
\left\{\begin{array}{l}
z_1^{\star}=0.09832 \\[0.3cm]
z_2^{\star}=0.09624 \\[0.3cm]
z_3^{\star}=0.09616 \\[0.3cm]
z_4^{\star}=0.08840
\end{array}\right.
=g_4\times \left\{\begin{array}{l}
36.28006186 \\[0.3cm]
35.51254224 \\[0.3cm]
35.48302226 \\[0.3cm]
32.61958369,
\end{array}\right.
\label{shiki49}
\end{equation}
where we have used (\ref{shiki10}). We also 
estimate the strength of each phase transition, which is 
given by
\begin{equation}
{v(T_{ic})\over T_{ic}}={1\over g_4}
{{\Delta a_0^{T=T_{c}}}\over {z_i^{\star}}}={1\over g_4}\times 
\left\{\begin{array}{ll}
0.1336339/z_1^{\star}=1.35922498 \\[0.3cm]
(0.262426-0.184866)/z_2^{\star}=0.80590191 \\[0.3cm]
(0.865538-0.271745)/z_3^{\star}=6.17505199 \\[0.3cm]
(0.79899-0.666667)/z_4^{\star}=1.49686652,
\end{array}\right.
\label{shiki50}
\end{equation}  
where $\Delta a_0^{T=T_{c}}\equiv
\abs{a_0^{T=T_{jc}}-a_0^{T=T_{kc}}}$
stands for the jump between the two VEV's at the 
critical temperature $z_i^{\star}(i=1\sim 4)$.
We see that except for the second phase transition, 
the phase transitions are the strong first order. 
It is interesting to consider the
application of the multi phase transition.
%%%%%%%%%%%%
\section{Conclusions}
We have studied the phase transition in gauge-Higgs unification 
at finite temperature in a certain class of gauge models, 
including supersymmetric ones. 
Since the dynamics of the gauge-Higgs unification is 
essentially the Coleman-Weinberg mechanism, 
the phase transition is strongly expected to be the first order.
We have, in fact, observed that the phase transition 
is the first order, and the strength becomes stronger as the
Higgs mass becomes lighter in models under consideration.
\par
%%%%%%%%%%%%  
We have also found an interesting phase transition patterns, 
which has never been reported in the theory of the gauge-Higgs 
unification for an $SU(3)$ gauge theory with only adjoint fermions. 
There have been two stage phase transitions characterized 
by the two critical temperatures $T_{1c}$ and $T_{2c}$. 
At each critical temperature, 
it is turned out that the phase transition is the first order. 
The vacua with the $SU(2)\times U(1)$ gauge symmetry and 
the $U(1)^{\prime}\times U(1)$ gauge symmetry 
are degenerate at $T=T_{c1}$, while at $T=T_{2c}(< T_{1c})$, 
the vacua with the $U(1)^{\prime}\times U(1)$ symmetry 
and the $U(1)_{em}$ gauge symmetry are degenerate. 
It is interesting to consider if there are applications of 
these two stage phase transitions. 
\par
%%%%%%%%%%%
We have also studied supersymmetric gauge models with 
the Scherk-Schwarz supersymmetry breaking. 
We have found that the phase transition is strongly 
the first order
for the simple matter content (\ref{shiki31}) 
which provides with the correct pattern of electroweak 
symmetry breaking 
and the Higgs mass of the correct order of magnitude. 
We have also studied the phase transition for
the matter content with supermultiplets only 
in the fundamental representation (\ref{shiki41}) and 
only in the adjoint representation (\ref{shiki46}).
We have found the same phase transition pattern as the case for 
non-SUSY gauge model with the corresponding matter content
(\ref{shiki17}) for the matter content (\ref{shiki41}).
On the other hand, for the matter content (\ref{shiki46}), 
we have obtained the multi phase (four in the case) transitions.
\par
%%%%%%%%% 
Contrary to the non-SUSY models, we have a free parameter 
in SUSY models, 
the SS parameter $\beta$ in addition to the number 
of flavors. 
In ref.\cite{hytmodel} it has been pointed out that the magnitude 
of $\beta$ affects 
the size of the Higgs mass, so that it is very 
interesting to study how 
the strength of the phase transition depends on 
the SS parameter. 
This will be reported in a separated paper.
%%%%%%%%%
Finally, in order to understand the phase structure of 
the theory of gauge-Higgs unification more deeply, 
it is necessary to study the effect of chemical 
potential $\mu$ \cite{actor}. If we introduce 
the chemical potential for a fermion, for example, 
the effective potential for the fermion is modified as 
\begin{eqnarray}
V_F^{\mu} &=& (-1)^F (r2^{[D/2]}) \half (-1) \sum_{n=1}^{\infty}
\Biggl[
{{2\Gamma(D/2)} \over{(2\pi R n)^D\pi^{D/2}}} \cos[2\pi n(Qa -
\half \delta)]
\nonumber\\
&-&
\frac{2 \pi^{D\over 2}}{ \Gamma(D/2)(2 \pi R)^D }
\left(n + Qa - \frac{\delta}{2} \right)^{D-1} 
\int_1^{\infty}dy~(y^2-1)^{{D \over 2}-1}
\left({1 \over{1 + \e^{{\bar z}y + {\mu \over T}}}} 
+ (\mu \leftrightarrow -\mu)
%
%{1 \over{1 + \e^{{\bar z}y - {\mu \over T}}}} 
%
\right)
\Biggr],\nonumber \\
\label{shiki51}
\end{eqnarray}
where ${\bar z}\equiv 2\pi(n+Qa-\delta/2)/2\pi R T,~Q=\half~(1)$ for 
matter belonging to the fundamental (adjoint) representations 
under the $SU(3)$ gauge group and $\delta$ 
takes $+1$ or $-1$, depending on the intrinsic parity. 
Since this potential is very complicated, 
we are particularly interested in the behavior of 
the effective potential in the limit of $T\rightarrow 0$. 
This work is in progress, and we hope that interesting 
results will be soon reported elsewhere. 
%%%%
%an interesting behavior and effect of the chemical
%potential on the gauge-Higgs unification. 
%%%%%%%%%%%%%%%%%%%%
\begin{center}
{\bf Acknowledgements}
\end{center}  
N.M. is supported by Special Postdoctral Researchers Program 
at RIKEN (No. A12-61014). 
K.T. would thank the professors, K. Funakubo and
Y. Hosotani for valuable discussions. K.T. is 
supported by the $21$st Century COE Program at Osaka University. 
%\vskip 2cm
%\newpage
%%%%%%%%%%%%% BIBLIOGRAPHY %%%%%%%%%%%%%%%%%%%%

%%%%%%%%%%%%%%%%%%%%%%%%%%%%
\end{document}